\documentclass[11pt,a4paper]{article}
\pdfoutput=1

\usepackage{jheppub}

\usepackage{graphicx}
\usepackage{feynmp-auto}
\usepackage{subcaption}
\usepackage{amsmath}
\usepackage{amssymb}
\usepackage{amsxtra}

\def\be{\begin{eqnarray}}
\def\ee{\end{eqnarray}}
\def\0{\nonumber}
\def\d{\partial}
\def\tr{{\rm tr}} 
\def\del{\partial}

\usepackage{slashed}
\usepackage{verbatim}
\usepackage{latexsym}
\usepackage{euscript}
\newcommand\ER{\EuScript{R}}

\newcommand\EW{\EuScript{W}}

\newcommand\ED{\EuScript{D}}\normalfont\large

\preprint{SISSA/09/2014/FISI\\{\tt hep-th/1403.2606}}

\title{Trace anomalies in chiral theories revisited}

\author[a,b]{Loriano Bonora,}
\author[c]{Stefano Giaccari}
\author[a]{and Bruno Lima de Souza}

\affiliation[a]{International School for Advanced Studies (SISSA),\\Via Bonomea 265, 34136 Trieste, Italy}
\affiliation[b]{INFN, Sezione di Trieste, Italy}
\affiliation[c]{Department of Physics \& Center for Field Theory and Particle Physics, Fudan University\\200433 Shanghai, China}

\emailAdd{bonora@sissa.it}
\emailAdd{giaccari@fudan.edu.cn}
\emailAdd{blima@sissa.it}
   
\abstract{Motivated by the search for possible CP violating terms in the trace of the energy-momentum tensor in theories coupled to gravity we revisit the problem of trace anomalies in chiral theories. We recalculate the latter and ascertain that in the trace of the energy-momentum tensor of theories with chiral fermions at one-loop the Pontryagin density appears with an imaginary coefficient. We argue that this may break unitarity, in which case the trace anomaly has to be used as a selective criterion for theories, analogous to the chiral anomalies in gauge theories. We analyze some remarkable consequences of this fact, that seem to have been overlooked in the literature.}
 

\begin{document}
\maketitle
\flushbottom

\section{Introduction}\label{s:intro}

We revisit trace anomalies in theories coupled to gravity, an old subject, \cite{capper,deser,bernard,brown,brown-cas,chris,adler,duff1,dowker,tsao,ChD,vilenkin,wald,CD2,Christensen:1978sc,Duff:1982yw,duff2}, brought back to people's attention thanks to the importance acquired recently by conformal field theories both in themselves and in relation to the AdS/CFT correspondence. What has stimulated specifically this research is the suggestion by \cite{Nakayama,Nakayama:2013oqa} that trace anomalies may contain a CP violating term (the Pontryagin density). It is well known that a basic condition for baryogenesis is the existence of CP non-conserving reactions in an early stage of the universe. Many possible mechanisms for this have been put forward, but to date none is completely satisfactory (see, for instance, \cite{dolgov}). The appearance of a CP violating term in the trace anomaly of a theory weakly coupled to gravity may provide a so far unexplored new mechanism for baryogenesis.

Let us recall that the energy-momentum tensor in field theory is defined by
$T_{\mu\nu}=\frac{2}{\sqrt{|g|}}\frac{\delta S}{\delta g^{\mu\nu}}$. Under an infinitesimal local rescaling of the metric: $\delta g_{\mu\nu}=2\sigma g_{\mu\nu}$ we have
\begin{equation}
\delta S=\frac{1}{2}\int d^{4}x\sqrt{|g|}T_{\mu\nu}\delta g^{\mu\nu}=-\int d^{4}x\sqrt{|g|}\sigma T_{\mu}^{\phantom{\mu}\mu}.\0
\end{equation}
If the action is invariant, classically $T_{\mu}^{\phantom{\mu}\mu}=0$, but at one-loop (in which case $S$ is replaced by the one-loop effective action $W$) the trace of the e.m. tensor is generically non-vanishing. 
In $D=4$ it may contain, \cite{BPT}, in principle, beside the Weyl density (square of the Weyl tensor)
\be
\EW^2=\ER_{nmkl} \ER^{nmkl}-2 \ER_{nm}\ER^{nm} +\frac 13 \ER^2\label{weyl}
\ee
and the Gauss-Bonnet (or Euler) one,
\be
E=\ER_{nmkl} \ER^{nmkl}-4 \ER_{nm}\ER^{nm} + \ER^2,\label{gausbonnet}
\ee
another nontrivial piece, the Pontryagin density,\footnote{The $\epsilon^{nmlk}$ figuring in (\ref{pontryagin}) is the Levi-Civita \emph{tensor}, which means that it is the usual Levi-Civita symbol divided by $\sqrt{|g|}$. We also recall that the Pontryagin density is a type-B anomaly according to the classification of \cite{Deser:1993yx}, see also \cite{Boulanger:2007st}.}
\be
P=\frac 12\left(\epsilon^{nmlk}\ER_{nmpq}\ER_{lk}{}^{pq}\right)\label{pontryagin}.
\ee
Each of these terms appears in the trace with its own coefficient:
\be
T_\mu{}^\mu= a E+ c \EW^2 +e P\label{emtrace}
\ee
The coefficient $a$ and $c$ are known at one-loop for any type of matter (see \cite{parker} for a review and the textbooks \cite{bert,fuji,BvN} for various techniques used). The coefficient of (\ref{pontryagin}) has not been sufficiently studied yet. The purpose of this paper is to contribute to fill this gap. More specifically we analyse the one-loop calculation of the trace anomaly in chiral models. Both the problem and the relevant results cannot be considered new, they are somehow implicit in the literature (see for example \cite{Duff:1982yw}): the trace anomaly contains beside the square Weyl density and the Euler density also the Pontryagin density. What is important, and we stress in this paper, is that the $e$ coefficient is purely imaginary. This may entail a problem of unitarity at one-loop. We argue that this introduces an additional consistency criterion for a theory. The latter has to be compared with the analogous criterion for chiral gauge and gravitational anomalies, which is since long a selective criterion for consistent theories. 
This may have important physical consequences, as will be pointed out in the conclusive sections.

In this paper, for simplicity, we will examine the problem of the one-loop trace anomaly in a prototype chiral theory, that of a free chiral fermion coupled to external gravity. In section \ref{sec:2} we calculate the CP violating part of the trace anomaly in such a model using the heat kernel and zeta function regularization, already available in the literature. In section \ref{sec:3} we do the same using Feynman diagram techniques. In section \ref{sec:4}, as an example, we apply these results to the standard model in its old and modern formulation. Finally in the last section we discuss some delicate aspects of the previous section. Three appendices are devoted to some details of the calculations carried out in section \ref{sec:3}.

\section{One-loop trace anomaly in chiral theories}
\label{sec:2}

The model we will consider is the simplest possible one: a right-handed spinor coupled to external gravity in $4d$. The action is
\be
S= \int d^4x \, \sqrt{|g|} \, i\overline {\psi_R} \gamma^m\left(\nabla_m +\frac 12 \omega_m \right)\psi_R \label{action},
\ee
where $\gamma^m = e^m_a \gamma^a$, $\nabla$ ($m,n,\dots$ are world indices, $a,b,\dots$ are flat indices) is the covariant derivative with respect to the world indices and $\omega_m$ is the spin connection:
\be
\omega_m= \omega_m^{ab} \Sigma_{ab}\0
\ee
where $\Sigma_{ab} = \frac 14 [\gamma_a,\gamma_b]$ are the Lorentz generators. Finally
$\psi_R= \frac {1+\gamma_5}2 \psi$. Classically the energy-momentum tensor 
\be
T_{\mu\nu}= \frac i4 \overline {\psi_R} \gamma_\mu {\stackrel{\leftrightarrow}{\nabla}}_\nu\psi_R+ \left(\mu\leftrightarrow \nu\right)
 \label{emt}
\ee
is both conserved and traceless on-shell. At one-loop to make sense of the calculations one must introduce regulators. The latter generally breaks both diffeomorphism and conformal invariance. A careful choice of the regularization procedure may preserve diff invariance, but anyhow breaks conformal invariance, so that the trace of the e.m. tensor takes the form (\ref{emtrace}), with specific non-vanishing coefficients $a$, $c$ and $e$. There are various techniques to calculate the latter: cutoff, point splitting, Pauli-Villars, dimensional regularization and etc. Here we would like to briefly recall the heat kernel method utilized in \cite{CD2} and in  references cited  therein. Denoting by $D$ the relevant Dirac operator in (\ref{action}) one can show that
\be 
\delta W =  -\int d^{4}x\sqrt{|g|}\sigma T_{\mu}^{\phantom{\mu}\mu}=-\frac{1}{16\pi^{2}}\int d^{4}x\sqrt{|g|}\sigma b_{4}\left(x,x;D^{\dagger}D\right).\0
\ee
Thus
\be
T_{\mu}^{\phantom{\mu}\mu}= \frac{1}{16\pi^{2}} b_{4}\left(x,x;D^{\dagger}D\right)\label{traceb4}
\ee
The coefficient $b_{4}\left(x,x;D^{\dagger}D\right)$ appears in the heat kernel. The latter has the general form
\begin{equation}
K\left(t,x,y;\ED\right)\sim\frac{1}{\left(4\pi t\right)^{2}}e^{-\frac{\sigma\left(x,y\right)}{2t}}\left(1+tb_{2}\left(x,y; \ED\right)+t^{2}b_{4}\left(x,y; \ED\right)+\cdots\right),\0
\end{equation}
where $\ED=D^\dagger D$ and $\sigma\left(x,y\right)$ is the half square length
of the geodesic connecting $x$ and $y$, so that $\sigma\left(x,x\right)=0$.
For coincident points we therefore have
\begin{equation}
K\left(t,x,x;\ED\right)\sim\frac{1}{16\pi^{2}}\left(\frac{1}{t^{2}}+\frac{1}{t}b_{2}\left(x,x;\ED\right)+b_{4}\left(x,x;\ED\right)+\cdots\right). 
\end{equation}
This expression is divergent for $t\to 0$ and needs to be regularized. This can be done in various ways. The finite part, which we are interested in, has been calculated first by DeWitt, \cite{DeWitt}, and then by others with different methods. The results are reported in \cite{CD2}. For a spin $\frac{1}{2}$ right-handed spinor as in our example one gets 
\begin{equation}
b_{4}\left(x,x;D^{\dagger}D\right)  =  \frac{1}{180}\left(a\, E_{4}+c\, W^{2}+e\, P\right),
\end{equation}
with
\be
a= \frac {11}{4},\quad\quad c= -\frac {9}{2},\quad\quad e= \frac {15}{4}\label{coefficients}.
\ee
This result was obtained with an entirely Euclidean calculation. Coming back to Lorentzian signature the e.m. trace at one-loop is
\be
T_\mu{}^\mu= \frac{1}{180\times16\pi^{2}}\left(\frac {11}{4} \,  E-\frac {9}{2}\, \EW^2 +i \frac{15}{4}\, P\right)\label{emtracechiral}.
\ee
As pointed out above the important aspect of (\ref{emtracechiral}) is the $i$ appearing in front of the Pontryagin density. The origin of this imaginary coupling is easy to trace. It comes from the trace of gamma matrices including a $\gamma_5$ factor. In $4d$, while the trace of an even number of gamma matrices, which give rise to first two terms in the RHS of (\ref{emtracechiral}), is  a real number, the trace of an even number of gamma's multiplied by $\gamma_5$ is always imaginary. The Pontryagin term comes precisely from the latter type of traces. It follows that, as a one loop effect, the energy momentum tensor becomes complex, and, in particular, since $T_0^0$ is the Hamiltonian density, we must conclude that unitarity may  not be preserved in this type of theories. It is legitimate to ask whether, much like chiral gauge theories with non-vanishing chiral gauge anomalies are rejected as sick theories, also chiral models with complex trace anomalies are not acceptable theories. We will return to this point later on.

\section{Other derivations of the Pontryagin trace anomaly}
\label{sec:3}

The derivation of the results in the previous section are essentially based on the method invented by DeWitt, \cite{DeWitt}, which is a point splitting method, the splitting distance being geodesic. As such, it guarantees covariance of the anomaly expression. To our surprise we have found that, while for the CP preserving part of the trace anomaly various methods of calculation are available in the literature, no other method is met to calculate the coefficient of the Pontryagin density. Given the important consequences of such a (imaginary) coefficient, we have decided to recalculate the results of the previous section with a different method, based on Feynman diagram techniques. We will use it in conjunction with dimensional regularization.

To start with from (\ref{action}) we have to extract the Feynman rules.\footnote{We follow closely the derivation of the chiral anomaly in \cite{DS}, although with a different regularization. For other derivations of this anomaly see also \cite{eguchi,endo}.}
More explicitly the action (\ref{action}) can be written as
\be 
S= \int d^4x \, \sqrt{|g|} \,\left[ \frac{i}{2}\overline {\psi_R} \gamma^\mu {\stackrel{\leftrightarrow}{\d}}_\mu  \psi_R -\frac{1}{4}\epsilon^{\mu a b c} \omega_{\mu a b} \overline{\psi_R} \gamma_c \gamma_5\psi_R\right]
\label{action1}
\ee
where it is understood that the derivative applies to $\psi_R$ and $\overline {\psi_R}$ only. We have used the relation $\{\gamma^a, \Sigma^{bc}\}=i \, \epsilon^{abcd}\gamma_d\gamma_5$.
Now we expand
\be
e_\mu^a = \delta_\mu^a +\chi_\mu^a+\cdots,\quad e_a^\mu = \delta _a^\mu +\hat \chi_a^\mu +\cdots,\quad \text{and}\quad
g_{\mu\nu}=\eta_{\mu\nu}+h_{\mu\nu}+\cdots\label{hmunu}
\ee
Inserting these expansions in the defining relations $e^a_\mu e^\mu_b=\delta_b^a$, $g_{\mu\nu}= e_\mu^a e_\nu^b \eta_{ab}$,
we find
\be
\hat \chi_\nu^\mu =- \chi_\nu^\mu\quad\text{and}\quad h_{\mu\nu}=2\,\chi_{\mu\nu}.\label{hatchichi}
\ee
From now on we will use both $\chi_\mu^a$ and $h_{\mu\nu}$, since we are interested in the lowest order contribution, we will raise and lower the indices {\it only} with $\delta$. We will not need to pay attention to the distinction between flat and world indices.
Let us expand accordingly the spin connection. Using
\be
\omega_{\mu ab}=e_{\nu a}( \d_\mu e_b^\nu + e^\sigma{}_b \Gamma_\sigma{}^\nu{}_\mu)\quad \text{and}\quad
\Gamma_\sigma{}^\nu{}_\mu= \frac{1}{2} \eta^{\nu\lambda}(\d_\sigma h_{\lambda\mu}+\d_\mu h_{\lambda\sigma}-\d_\lambda
h_{\sigma\mu})+\cdots,\0
\ee
after some algebra we get
\be 
\omega_{\mu a b}\, \epsilon^{\mu a b c}= - \epsilon^{\mu a b c}\,  \d_\mu \chi_{a\lambda}\,\chi_b^\lambda+\cdots\label{omega}.
\ee
For later use let us quote the following approximation for the Pontryagin density
\be
\epsilon^{\mu\nu\lambda \rho}R_{\mu\nu}{}^{\sigma\tau}R_{\lambda\rho\sigma\tau}=
8 \epsilon^{\mu\nu\lambda \rho} \left(\d_\mu\d_\sigma \chi^a_\nu \, \d_\lambda\d_a \chi_{\rho}^\sigma-\d_\mu\d_\sigma \chi^a_\nu \, \d_\lambda\d^\sigma \chi_{a\rho}\right)+\cdots\label{epsRR}
\ee 
Therefore, up to second order the action can be written (by incorporating $\sqrt{|g|}$ in a redefinition of the $\psi$ field \footnote{This is the simplest way to deal with  $\sqrt{|g|}$. Alternatively one can keep it explicitly in the action and approximate it as $1 +\frac{1}{2} h^\mu_\mu$; this would produce two additional vertices, which however do not contribute to our  final result.})
\be 
S\approx \int d^4x \,  \left[\frac{i}{2} (\delta^\mu_a -\chi^\mu_a ) \overline {\psi_R} \gamma^a {\stackrel{\leftrightarrow}{\d}}_\mu  \psi_R +\frac{1}{4}\epsilon^{\mu a b c}\,  \d_\mu \chi_{a\lambda}\,\chi_b^\lambda\,  \overline\psi_R \gamma_c \gamma_5\psi_R\right]\0
\ee
The free action is 
\be
S_{free}=  \int d^4x \,\frac{i}{2} \overline {\psi_R} \gamma^a {\stackrel{\leftrightarrow}{\d}}_a  \psi_R\label{free}
\ee
and the lowest interaction terms are
\be
S_{int} &=&  \int d^4x \left[ -\frac{i}{2} \chi^\mu_a\,\overline {\psi_R} \gamma^a {\stackrel{\leftrightarrow}{\d}}_\mu  \psi_R+\frac{1}{4} \epsilon^{\mu a b c}\,  \d_\mu \chi_{a\lambda}\,\chi_b^\lambda\,  \overline\psi_R \gamma_c \gamma_5\psi_R\right]\0\\
&=& \int d^4x \left[ -\frac{i}{4} h^\mu_a\,\overline {\psi_R}\gamma^a {\stackrel{\leftrightarrow}{\d}}_\mu  \psi_R+\frac{1}{16} \epsilon^{\mu a b c}\,  \d_\mu h_{a\lambda}\,h_b^\lambda\,  \overline\psi_R \gamma_c \gamma_5\psi_R\right]\label{int}
\ee
As a consequence of (\ref{free}) and (\ref{int}) the Feynman rules are as follows (the external gravitational field is assumed to be $h_{\mu\nu}$). The fermion propagator is
\unitlength = 1mm
\begin{equation}
\parbox{20mm}{\begin{fmffile}{propagator}
	\begin{fmfgraph*}(20,20)
		\fmfleft{i}\fmfright{f}
		\fmf{fermion,tension=0,label=$p$}{i,f}
	\end{fmfgraph*}
\end{fmffile}}  =   \dfrac{i}{\slashed{p}+i\epsilon}.
\label{prop}
\end{equation}
The two-fermion-one-graviton vertex ($V_{ffg}$) is
\begin{equation}
\parbox{20mm}{\begin{fmffile}{Vffg}
	\begin{fmfgraph*}(20,20)
		\fmfleft{i}\fmfright{f1,f2}
		\fmf{photon}{i,v}
		\fmf{fermion,label=$p'$,l.side=right}{v,f1}\
		\fmf{fermion,label=$p$,l.side=right}{f2,v}
	\end{fmfgraph*}
\end{fmffile}}  =  \dfrac{i}{8}\left[\left(p+p'\right)_\mu\gamma_\nu+\left(p+p'\right)_\nu\gamma_\mu\right]\frac{1+\gamma_5}{2}.
\label{2f1g}
\end{equation}
The two-fermion-two-graviton vertex ($V_{ffgg}$) is
\begin{equation}
\parbox{20mm}{\begin{fmffile}{Vffgg}
	\begin{fmfgraph*}(20,20)
		\fmfleft{i1,i2}\fmfright{f1,f2}
		\fmf{photon,label=$k$,l.side=right}{i1,v}
		\fmf{photon,label=$k'$,l.side=right}{i2,v}
		\fmf{fermion,label=$p'$,l.side=left}{v,f1}\
		\fmf{fermion,label=$p$,l.side=right}{f2,v}
	\end{fmfgraph*}
\end{fmffile}}  =  \dfrac{1}{64}t_{\mu\nu\mu'\nu'\kappa\lambda}\left(k-k'\right)^\lambda\gamma^\kappa\dfrac{1+\gamma_5}{2},
\label{2f2g}
\end{equation}
where the momenta of the gravitons are ingoing and
\be
t_{\mu\nu\mu'\nu'\kappa\lambda}=\eta_{\mu\mu'} \epsilon_{\nu\nu'\kappa\lambda} +\eta_{\nu\nu'} \epsilon_{\mu\mu'\kappa\lambda} +\eta_{\mu\nu'} \epsilon_{\nu\mu'\kappa\lambda} +\eta_{\nu\mu'} \epsilon_{\mu\nu'\kappa\lambda}\label{t}.
\ee
Due to the non-polynomial character of the action the diagrams contributing to the trace anomaly are infinitely many. Fortunately, using diffeomorphism invariance, it is enough to determine the lowest order contributions and consistency takes care of the rest. There are two potential lowest order diagrams (see figures \ref{DiagramD} and \ref{TriangleDiagrams} in the appendices \ref{sec:C1} and \ref{sec:C2}) that may contribute. The first contribution, the bubble graph, turns out to vanish, see appendix \ref{sec:C1}. It remains for us to calculate the triangle graph. To limit the size and number of formulas in the sequel {\it we will be concerned only with the contribution of the diagrams to the Pontryagin density.}

\subsection{The fermion triangle diagram}
\label{ssec:3.2}

It is constructed by joining three vertices $V_{ffg}$ with three fermion lines. The external momenta are $q$ (ingoing) with labels $\sigma$ and $\tau$, and $k_1,k_2$ (outgoing), with labels
$\mu,\nu$ and $\mu',\nu'$ respectively. Of course $q=k_1+k_2$.
The internal momenta are $p $, $p-k_1$ and 
$p-k_1-k_2$, respectively. After contracting $\sigma$ and $\tau$ the total contribution is
\begin{align} 
&-\frac{1}{256}\int \frac{d^4p}{(2\pi)^4}\, {\rm tr} \left[\left(\frac{1}{\slashed{p}}\left((2p-k_1)_\mu \gamma_\nu+(\mu\leftrightarrow \nu)\right)\right.\right.\frac{1}{\slashed{p}-\slashed{k}_1} \nonumber\\
&\times \left((2p-2 k_1 - k_2)_{\mu'}\gamma_{\nu'}+(\mu'\leftrightarrow \nu')\right) \left.\left.\frac{1}{\slashed{p} - \slashed{k}_1 -\slashed{k}_2} (2\slashed{p} -\slashed{k}_1 -\slashed{k}_2)\right) \frac{1+\gamma_5}{2}\right] \label{T1}
\end{align}
to which we have to add the cross diagram in which $k_1,\mu,\nu$ is exchanged with $k_2,\mu',\nu'$. This integral is divergent. To regularize it we use dimensional regularization. To this end we introduce additional components of the momentum running on the 
loop (for details see, for instance, \cite{bert}): $p\to p+\ell$, $\ell=(\ell_4,\ldots, \ell_{n-4})$
\begin{align}
&T_{\mu\nu\mu'\nu'}(k_1,k_2)
= -\frac{1}{256} \int \frac {d^4p}{(2\pi)^4}
\int \frac {d^{n-4}\ell}{(2\pi)^{n-4}} \,\tr\left( \frac {\slashed{p}+\slashed{\ell}}{p^2-\ell^2} (2p+2\ell-k_1)_\mu\gamma_\nu \,\right.\0\\
&\times \,\frac {\slashed{p}+\slashed{\ell}-\slashed{k}_1}{(p-k_1)^2-\ell^2} \,(2p+2\ell-2k_1-k_2)_{\mu'}\gamma_{\nu'} \left.\frac {\slashed{p}+\slashed{\ell}-\slashed{q}}{(p-q)^2-\ell^2}\,( 2\slashed{p}+2\slashed{\ell}-\slashed{q})\frac {1+\gamma_5}{2}\right)\label{T2}
\end{align}
where the symmetrization over $\mu,\nu$ and $\mu',\nu'$ has been understood.\footnote{Some attention has to be paid in introducing the additional momentum components $\ell$. Due to the chiral projectors in the $V_{ffg}$ vertex it would seem that $\slashed{\ell}$ should not be present in the first and third terms in (\ref{T2}) (because $[ \slashed{\ell},\gamma_5]=0$); however this regularization prescription would give a wrong result for the CP even part of the anomaly. The right prescription is (\ref{T2}).} After some manipulations this becomes
\begin{align}
T_{\mu\nu\mu'\nu'}(k_1,k_2)=\; & T^{(1)}_{\mu\nu\mu'\nu'}(k_1,k_2)+T^{(2)}_{\mu\nu\mu'\nu'}(k_1,k_2) \0\label{T3}\\
&-\frac{1}{256} \int \frac {d^4p}{(2\pi)^4}\int \frac {d^{n-4}\ell}{(2\pi)^{n-4}}\tr\left( \frac {\slashed{p}+\slashed{\ell}}{p^2-\ell^2} (2p+2\ell-k_1)_\mu\gamma_\nu \right.\0 \\
&\left.\times\frac {\slashed{p}+\slashed{\ell}-\slashed{k}_1}{(p-k_1)^2-\ell^2} \,(2p+2\ell-2k_1-k_2)_{\mu'}\gamma_{\nu'} \frac {\slashed{p}+\slashed{\ell}-\slashed{q}}{(p-q)^2-\ell^2} \, \slashed{\ell}\gamma_5\right)
\end{align}
The terms $T^{(1)},T^{(2)}$ turn out to vanish. The rest, after a Wick rotation (see appendix \ref{sec:C2}), gives
\be
T_{\mu\nu\mu'\nu'}(k_1,k_2)= \frac{1}{6144\pi^2} \left(k_1\cdot k_2  \,t_{\mu\nu\mu'\nu'\lambda\rho}\,- t^{(21)}_{\mu\nu\mu'\nu'\lambda\rho} \right)
k_1^\lambda k_2^\rho\, \label{T4}
\ee
where 
\be 
t^{(21)}_{\mu\nu\mu'\nu'\kappa\lambda}=k_{2\mu}k_{1\mu'} \epsilon_{\nu\nu'\kappa\lambda} + k_{2\nu}k_{1\nu'}\epsilon_{\mu\mu'\kappa\lambda} +k_{2\mu}k_{1\nu'} \epsilon_{\nu\mu'\kappa\lambda} +k_{2\nu}k_{1\mu'} \epsilon_{\mu\nu'\kappa\lambda}\label{t21}
\ee
Finally we have to add the cross graph contribution, obtained by $k_1,\mu,\nu \leftrightarrow k_2,\mu',\nu'$. Under this exchange the $t$ tensors transform as follows:
\be
t\leftrightarrow -t, \quad\quad t^{(21)} \leftrightarrow -t^{(21)}, \quad i\neq j\label{ttransf}
\ee
Therefore the cross graph gives the same contribution as (\ref{T4}). So for the triangle
diagram we get
\be
T^{(tot)}_{\mu\nu\mu'\nu'}(k_1,k_2)= \frac{1}{3072\pi^2} \left(k_1\cdot k_2  \,t_{\mu\nu\mu'\nu'\lambda\rho}\,- t^{(21)}_{\mu\nu\mu'\nu'\lambda\rho} \right)
k_1^\lambda k_2^\rho\, \label{Ttot}
\ee
To obtain the above results we have set the external lines on-shell. This deserves a comment.

\subsection{On-shell conditions}

Putting the external lines on-shell means that the corresponding fields have to satisfy the EOM of gravity $R_{\mu\nu}=0$.  In the linearized form this means
\be 
\square \chi_{\mu\nu}= \d_{\mu} \d_{\lambda}\chi^\lambda _\nu + \d_\nu \d_{\lambda}\chi^\lambda _\mu -\d_\mu\d_\nu \chi^\lambda_\lambda\label{linrmunu}
\ee
We also choose the de Donder gauge
\be 
\Gamma_{\mu\nu}^\lambda g^{\mu\nu}=0\label{dedonder}
\ee
which at the linearized level becomes
\be
2\d_\mu\chi^{\mu}_\lambda -\d_\lambda \chi^\mu_\mu=0\label{lindedonder}
\ee
In this gauge (\ref{linrmunu}) becomes
\be 
\square \chi_{\mu\nu}= 0\label{KG}
\ee
In momentum space this implies that $k_1^2=k_2^2=0$. Since we know that the final result is covariant this simplification does not jeopardize it.

\subsection{Overall contribution}
\label{ssec:3.3}

The overall one-loop contribution to the trace anomaly in momentum space, {\it as far as the CP violating part is concerned}, is given by (\ref{Ttot}). After returning to the Minkowski metric 
and Fourier-antitransforming this we can extract the local expression of the trace anomaly, as explained in appendix \ref{sec:D}. The saturation with $h^{\mu\nu}, h^{\mu'\nu'}$ brings a 
multiplication by 4 of the anomaly coefficient. The result is, to lowest order,
\be 
\langle T^{\mu}_{\mu}(x)\rangle = \frac{i}{768\pi^2}\epsilon^{\mu\nu\lambda \rho} \left(\d_\mu\d_\sigma h^\tau_\nu \, \d_\lambda\d_\tau h_{\rho}^\sigma-\d_\mu\d_\sigma h^\tau_\nu \, \d_\lambda\d^\sigma h_{\tau\rho}\right)\label{final1}
\ee
Comparing with (\ref{epsRR}) we deduce the covariant expression of the CP violating part of the trace anomaly
\be 
\langle T^{\mu}_{\mu}(x)\rangle = \frac{i}{768\pi^2} \, \frac{1}{2}\,\epsilon^{\mu\nu\lambda \rho}R_{\mu\nu}{}^{\sigma\tau}R_{\lambda\rho\sigma\tau}\label{final2}
\ee
which is the same as (\ref{emtracechiral}).

\section{Consequences of the Pontryagin trace anomaly in chiral theories}
\label{sec:4}

In this section we would like to expand on the consequences of a non-vanishing Pontryagin term
in the trace anomaly. To start with let us spend a few words on a misconception we sometime come across: the gravitational charge of matter is its mass and, as a consequence, gravity interacts with matter via its mass. This would imply in particular that massless particles do not feel gravity, which is clearly false (e.g., the photon). The point is that gravity interacts with matter via its energy-momentum tensor. In particular, for what concerns us here, the e.m. tensor is different for left-handed and right-handed massless matter, and this is the origin of a different trace anomaly for them. 

As we have already noticed in \ref{sec:2}, in theories with a chiral unbalance, as a consequence of the Pontryagin trace anomaly, the energy momentum tensor becomes complex, and, in particular, unitarity is not preserved. This raises a question: much like chiral gauge theories with non-vanishing chiral gauge anomalies are rejected as unfit theories, should we conclude also that chiral models with complex trace anomalies are not acceptable theories? To answer this question it is important to put it in the right framework. To start with
let us consider the example of the standard model. In its pre-neutrino-mass-discovery period its spectrum was usually written as follows:
\be
\left(\begin{matrix} u\\ d\end{matrix}\right)_L, \quad \widehat {u_R},\quad \widehat{d_R},
\quad   \left(\begin{matrix} \nu_e\\ e\end{matrix}\right)_L,\quad  \widehat{e_R}\label{SM}
\ee
together with two analogous families (here and in the sequel, for any fermion field $\psi$,   $\hat \psi= \gamma^0 C\psi^*$, where $C$ is the charge conjugation matrix, i.e. $\hat \psi$ represents the Lorentz covariant conjugate field). All the fields are Weyl spinors and a hat represents CP conjugation. If a
field is right-handed its CP conjugate is left-handed. Thus all the fields in (\ref{SM}) are left-handed. This is the well-known chiral formulation of the SM. So we could represent the entire family as a unique left-handed spinor $\psi_L$ and  write the kinetic part of the action as in (\ref{action}). However the coupling to gravity of a CP conjugate field is better described as follows (see, for instance, \cite{pal}). First, for a generic spinor field $\psi$, let us define
(with $L=\frac {1-\gamma_5}{2}, R= \frac {1+\gamma_5}{2}$, and $\psi_L=L\psi, \psi_R=R\psi$)
\be
\widehat {\psi_R}= \gamma^0 C \psi_R^*= \gamma^0 C R^* \psi^*=L \gamma^0 C\psi^*=L\hat \psi =
\hat {\psi}_L\label{psiLpsiR}
\ee
where we have used the properties of the gamma matrices and the charge conjugation matrix $C$:
\be 
C^{-1}\gamma_\mu C=-\gamma_\mu^T,\quad\quad CC^\dagger=1,\quad\quad CC^*=-1,\quad\quad
C^T=-C\0
\ee
and in particular $C^{-1}\gamma_5 C=\gamma_5^T$. Let us stress in (\ref{psiLpsiR}) the difference implied by the use of  $\,\,\widehat{}\,$ and  $\,\,\hat{}\,$, respectively.

With the help of these properties one can easily show that
\be
&&\sqrt{|g|}\, \overline {{\hat\psi}_L} \,\gamma^m\left(\nabla_m +\frac{1}{2} \omega_m \right)\hat\psi_L=
\sqrt{|g|} \, \overline{\widehat{\psi_R}}\, \gamma^m\left(\nabla_m +\frac{1}{2} \omega_m \right)\widehat{\psi_R}\0\\
&& =\sqrt{|g|}\,\psi_R^T\, C^{-1} \gamma^m \gamma^0 \left(\nabla_m +\frac{1}{2} \omega_m \right) C \psi_L^*\0
\ee
which, after a partial integration and an overall transposition, becomes
\be
\sqrt{|g|}\, \overline {\psi_R}\, \gamma^m\left(\nabla_m +\frac{1}{2} \omega_m \right)\psi_R\label{rightaction}
\ee
i.e. the right-handed companion of the initial left-handed action. This follows in particular from 
the property $C^{-1} \Sigma_{ab}C= - \Sigma_{ab}^T$. 

From the above we see that in the multiplet (\ref{SM}) there is a balance between the left-handed and right-handed field components except for the left-handed field $\nu_e$. Therefore
the multiplet (\ref{SM}) when weakly coupled to gravity, will produce an overall non-vanishing (imaginary) coefficient $e$ for the Pontryagin density in the trace anomaly and, in general, a breakdown of unitarity (this argument must be seen in the context of the discussion in the following section). This breakdown is naturally avoided if we add to the SM multiplet a right-handed neutrino field, because in that case the balance of chirality is perfect.
Another possibility is that the unique neutrino field in the multiplet be Majorana, because a Majorana fermion can be viewed as a superposition of a left-handed and a right-handed Weyl spinor, with the additional condition of reality, and, therefore its contribution to the Pontryagin density is null. In both cases the neutrino can have mass. 

In hindsight this could have been an argument in favor of massive neutrinos.

From a certain point of view what we have just said may sound puzzling because it is often stated that in 4D massless Weyl and Majorana fermions are physically indistinguishable: they have the same number of components and we can define a one-to-one correspondence between the latter. A theory of Majorana fermions cannot have the kind of (chiral) anomaly we have found. So where does our anomaly comes from? It is therefore necessary to spend some time recalling the crucial difference between Weyl and Majorana fields in 4D. To start with, the map between Majorana and Weyl fields mentioned above is not representable by means of a linear invertible operator and this fact radically changes the way they transform under Lorentz transformations. Majorana fields transform as real representations and Weyl fields as complex representations of the Lorentz group. As a consequence, the relevant Dirac operators are different. Now, when we compute anomalies using the path integral we have to integrate  over fields, not over particles. Therefore anomalies are determined by the field content of the theory and by the appropriate Dirac operator. On the other hand anomalies like our Pontryagin anomaly (and many others) are not physical objects, but defects of the theory. Thus what we 
are saying is: if we want to formulate a theory with a different number of left-handed and right-handed Weyl fields, we are bound to find a dangerous anomaly in the trace of the em tensor. This does not prevent us from constructing a theory with the same physical content in an another way, which is anomaly-free, by using Majorana fields. But the path integrals of the two theories are not coincident. This, in turn, is connected with a related question: it is well known that, by means of mere algebraic manipulations, we can rewrite the kinetic action term of a Weyl field as the kinetic term of the corresponding Majorana field. So at first sight that seems to be no difference between the two. But this conclusion would forget that the transformation from Weyl to Majorana fields is not linear and invertible, so that one must take into account the Jacobian in the path integral. This is hard to compute directly, but what we have stressed in this paper is that it manifests itself (at least) in the Pontryagin anomaly.

\section{Discussion and conclusion}

The main point of this paper is a reassessment of the role of trace anomalies in theories with chiral matter coupled to gravity. In particular we have explicitly calculated the trace anomaly for a chiral fermion. The result is the expected one on the basis of the existing literature, except for the fact that, in our opinion, it had never been explicitly stated before (save for a footnote in  \cite{Nakayama,Nakayama:2013oqa}), and, especially, its consequences had never been seriously considered. As we have seen, for chiral matter the trace anomaly at one-loop contains the Pontryagin density $P$ with an imaginary coefficient. This implies, in particular, that the Hamiltonian density becomes complex and breaks unitarity. This poses the problem of whether this anomaly is on the same footing as chiral gauge anomalies in a chiral theory, which, when present, spoil its consistency. It is rightly stressed that the standard model is free of any chiral anomaly, including the gravitational ones. But in the case of ordinary chiral gauge anomalies the gauge fields propagate and drag the inconsistency in the internal loops, while in gravitational anomalies (including our trace anomaly) gravity is treated as a background field. 
So, do the latter have the same status as chiral gauge anomalies?

Let us analyse the question by asking: are there cases  
in which the Pontryagin density vanishes identically? The answer is: yes, there are background geometries where the Pontryagin density
vanishes. They include for instance the FRW and Schwarzschild \cite{Alexander:2009tp}. Therefore, in such backgrounds the problem of unitarity simply does not exist. But the previous ones are very special `macroscopic' geometries. For a generic geometry the Pontryagin density does not vanish. For instance in a cosmological framework, we can imagine to go up to higher energies where gravity inevitably back-reacts. In this case it does not seem to be possible to avoid the conclusion that the Pontryagin density does not vanish and unitarity is affected due to the trace anomaly, the more so because gravitational loops cannot cancel it. Thus, seen in this more general context, the breakdown of unitarity due to a chirality unbalance in an asymptotically free matter theory should be seriously taken into account.

Returning now to the problem we started with in the introduction, that is the appearance of a CP violating Pontryagin density in the trace of the energy-momentum tensor, we conclude that unitarity seems to prevent it at one-loop, and we cannot imagine a mechanism that may produce it at higher loops. In \cite{Nakayama,Nakayama:2013oqa} a holographic model was presented which yields a Pontryagin density in the trace of the e.m. tensor, but again with a unitarity problem \cite{Nakayama:2013oqa}. Anyhow it would be helpful to understand its (very likely, non-perturbative) origin in the boundary theory. This mechanism for CP violation is very interesting and, above, we have seen another attractive aspect of it: its effect evaporates automatically while the universe evolves towards `simpler' geometries.

A final comment about supersymmetry. In a previous paper, \cite{BG}, the compatibility between the appearance of the Pontryagin term in the trace anomaly and supersymmetry was considered and evidence was produced that they are not compatible. Altogether this and the results of this paper point towards the need for a theory which is neither chiral nor supersymmetric, if we wish to see the Pontryagin density with a real coefficient appear in the trace of the energy-momentum tensor. How this may actually be realized, as suggested in \cite{Nakayama,Nakayama:2013oqa}, is still an open and intriguing problem.

\acknowledgments 

We would like to thank Sergey Petcov and Roberto Percacci for useful discussions. We would like to thank Himanshu Raj and Marco Sanchioni for their collaboration in the early stage of this work. A short preliminary version of this paper will appear in the proceedings of the workshops ``Lie Theory and Its Applications in Physics", (LT-10),  17 - 23 June 2013, Varna (Bulgaria) and ``What comes beyond the standard model'' 14-21 July 2013 at Bled (Slovenia). L.B. would like to thank Norma Mankoc and Vladimir Dobrev for their invitation.

\vskip 1cm 
\section*{Appendix}
\appendix

\section{Calculation details: the bubble diagram}
\label{sec:C1}

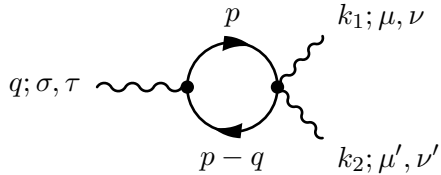
\begin{figure}[h]
\centering
\begin{fmffile}{em_tensor2}
	\begin{fmfgraph*}(30,40)
		\fmfleft{i} \fmflabel{$q;\sigma,\tau$}{i}
		\fmfright{fp1,f1,f2,fp2} \fmflabel{$k_2;\mu',\nu'$}{f1} \fmflabel{$k_1;\mu,\nu$}{f2}
		\fmf{photon}{i,v1}
		\fmf{photon}{v2,f1}
		\fmf{photon}{v2,f2}
		\fmf{fermion,left,tension=.5,label=$p$}{v1,v2}
		\fmf{fermion,left,tension=.5,label=$p-q$}{v2,v1}
		\fmfdot{v1,v2}
	\end{fmfgraph*}
\end{fmffile}
\caption{Bubble diagram with ingoing momentum $q$ and outgoing $k_1$ and $k_2$.}
\label{DiagramD}
\end{figure}

In this appendix we give a few details of the calculations in section \ref{sec:3}. 
Let us consider first the bubble graph (see figure \ref{DiagramD}).
It is obtained by joining two vertices, $V_{ffg}$ (on the left) and $V_{ffgg}$ (on the right) with two fermion propagators. The ingoing graviton in $V_{ffg}$ has momentum $q$ and Lorentz labels $\sigma,\tau$ and the two outgoing gravitons in $V_{ffgg}$ are specified by $k_1,\mu,\nu$ and $k_2,\mu',\nu'$, respectively. Of course $q=k_1+k_2$. 
The two fermion propagators form a loop. The running momentum is clockwise oriented. We denote 
the momentum in the upper branch of the loop by $p $ and in the lower branch by $p -q$.  This diagram is
\be 
2\times\frac{i}{512} \int \frac {d^4p}{(2\pi)^4}\, \tr \left[ \frac 1{\slashed{p} }\,t_{\mu\nu\mu'\nu'\lambda\rho}\,(k_2-k_1)^\lambda \gamma^\rho \frac 1{\slashed{p}-\slashed{q}} \left( \left(2p^\sigma - q^\sigma\right)\gamma^\tau + (\sigma \leftrightarrow \tau)\right) \frac {1+\gamma_5}2 \right]\quad\label{D1}
\ee 
The factor of two in front of it comes from the combinatorics of diagrams: this one must contributes twice. Its possible contribution to the trace anomaly comes from contracting the indices $\sigma$ and $\tau$ with a Kronecker delta (in principle we should consider contracting also the other couple of indices $\mu,\nu$ and $\mu',\nu'$, but this gives zero due to the symmetry properties of the $t$ tensor). The integral is divergent and needs to be regularized. We use dimensional regularization. To this end we introduce additional components of the momentum running on the 
loop: $p\to p+\ell$, $\ell=(\ell_4,\ldots, \ell_{n-4})$. The relevant integral becomes
\begin{align} 
D_{\mu\nu\mu'\nu'}(k_1,k_2)=\,& \frac i{256}\int \frac {d^4p}{(2\pi)^4}
\int \frac {d^{n-4}\ell}{(2\pi)^{n-4}} \, t_{\mu\nu\mu'\nu'\lambda\rho}(k_2-k_1)^\lambda\label{Dmumu}\0\\
&\times \,\tr \left(\frac {\slashed{p}+\slashed{\ell}}{p^2-\ell^2}\gamma^\rho \frac {\slashed{p}-\slashed{q}+\slashed{\ell}}{(p-q)^2-\ell^2} (2\slashed{p}+2\slashed{\ell}-\slashed{q})\right)
\end{align}
After some algebra and introducing a parametric representation for the denominators, one finally gets
\begin{align}
&D_{\mu\nu\mu'\nu'}(k_1,k_2)=- \frac i{64}\,t_{\mu\nu\mu'\nu'\lambda\rho}(k_2-k_1)^\lambda \int_0^1dx\int\frac {d^4p}{(2\pi)^4} \int \frac {d^{n-4}\ell}{(2\pi)^{n-4}}\0 \label{Dmumu2}\\
&\times \left[\left(\frac 32 (2x-1)p^2+x(x-1)(2x-1)q^2- (2x-1)\ell^2\right) q^\rho\right] \frac 1{(p^2+x(1-x)q^2-\ell^2)^2}
\end{align}
This vanishes because of the $x$ integration.

\section{Calculation details: the triangle diagram}
\label{sec:C2}

\begin{figure}[h]
\centering
\begin{subfigure}{.5\textwidth}
  \centering
  \begin{fmffile}{em_tensor1}
	\begin{fmfgraph*}(30,40)
		\fmfleft{i1,i2,i3} \fmflabel{$q;\sigma,\tau$}{i2}
    \fmfright{fp1,f1,f2,fp2} \fmflabel{$k_2;\mu',\nu'$}{f1} \fmflabel{$k_1;\mu,\nu$}{f2}
		\fmf{photon,tension=2.5}{i2,v1}
		\fmf{photon,tension=1}{v2,f1}
		\fmf{photon,tension=1}{v3,f2}
		\fmf{phantom}{v2,fp1}
		\fmf{phantom}{v3,fp2}
		\fmf{fermion,tension=1,label=$p$,l.side=left}{v1,v3}
		\fmf{fermion,tension=0.5,label=$p-k_1$,l.side=left}{v3,v2}
		\fmf{fermion,tension=1,label=$p-k_1-k_2$}{v2,v1}
		\fmfdot{v1,v2,v3}
	\end{fmfgraph*}
\end{fmffile}
\caption{Triangle diagram.}
\label{DiagramT}
\end{subfigure}%
\begin{subfigure}{.5\textwidth}
  \centering
  \begin{fmffile}{em_tensor1cross}
	\begin{fmfgraph*}(30,40)
		\fmfleft{i1,i2,i3} \fmflabel{$q;\sigma,\tau$}{i2}
    \fmfright{fp1,f1,f2,fp2} \fmflabel{$k_2;\mu',\nu'$}{f1} \fmflabel{$k_1;\mu,\nu$}{f2}
		\fmf{photon,tension=2}{i2,v1}
		\fmf{photon,tension=0}{v2,f2}
		\fmf{photon,tension=0}{v3,f1}
		\fmf{phantom}{v2,f1}
		\fmf{phantom}{v3,f2}
		\fmf{phantom}{v2,fp1}
		\fmf{phantom}{v3,fp2}
		\fmf{fermion,tension=1}{v1,v3}
		\fmf{fermion,tension=0.5}{v3,v2}
		\fmf{fermion,tension=1}{v2,v1}
		\fmfdot{v1,v2,v3}
	\end{fmfgraph*}
\end{fmffile}
\caption{Crossed triangle diagram.}
\label{DiagramTcross}
\end{subfigure}
\caption{In both these diagrams the momentum $q$ is ingoing while the momenta $k_1$ and $k_2$ is outgoing.}
\label{TriangleDiagrams}
\end{figure}
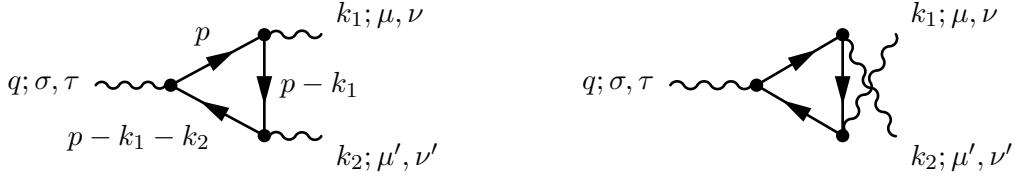	
As for the triangle diagram (see figure \ref{TriangleDiagrams}), with reference to eq.(\ref{T3}), we have
\begin{align}
T^{(1)}_{\mu\nu\mu'\nu'}(k_1,k_2) =&
-\frac 1 {256} \int \frac {d^4p}{(2\pi)^4}\int \frac {d^{n-4}\ell}{(2\pi)^{n-4}}\tr\left( \frac {\slashed{p}+\slashed{\ell}}{p^2-\ell^2} (2p+2\ell-k_1)_\mu\gamma_\nu\right.\label{C1}\0 \\
&\times \left.\frac {\slashed{p}+\slashed{\ell}-\slashed{k}_1}{(p-k_1)^2-\ell^2} \,(2p+2\ell-2k_1-k_2)_{\mu'}\gamma_{\nu'} \frac  {\gamma_5}2\right)\0\\
=& -\frac i {256}\int_0^1dx \int \frac {d^4p}{(2\pi)^4}
\int \frac {d^{n-4}\ell}{(2\pi)^{n-4}} \epsilon_{\nu \nu'\lambda\rho} k_1^\rho \frac {p^2}{(p^2+x(1-x)k_1^2-\ell^2)^2}\0\\
&\times \left( \delta_\mu^\lambda (2\ell-2xk_1-k_2)_{\mu'}+ \delta_{\mu'}^\lambda (2\ell-2xk_1-k_2)_{\mu}\right),
\end{align}
which evidently vanishes when we symmetrize $\mu$ with $\nu$ and $\mu'$ with $\nu'$.
$T^{(2)}$ is similar to $T^{(1)}$ and vanishes for the same reason.
Setting $k_1^2=k_2^2=0$, the remaining term in (\ref{T3}) can be written
\be 
T_{\mu\nu\mu'\nu'}(k_1,k_2) &=&\frac i{32}  \epsilon_{\nu \nu'\lambda\rho}k_1^\lambda k_2^\rho
\int_0^1dx \int_0^{1-x}dy \int \frac {d^4p}{(2\pi)^4}
\int \frac {d^{n-4}\ell}{(2\pi)^{n-4}} \0\label{C2}\\
&&\times\, \ell^2\, \frac {p^2 \eta_{\mu\mu'}+((2x+2y-1)k_1+2yk_2)_\mu(2(x+y-1)k_1+(2y-1)k_2)_{\mu'}}
{(p^2-\ell^2+2y(1-x-y)k_1\!\cdot \!k_2)^3}.\0\\
\ee
It involves two integrals over the momenta
\be 
\int\frac {d^4p}{(2\pi)^4}
\int \frac {d^{n-4}\ell}{(2\pi)^{n-4}}  \frac {\ell^2}{(p^2+x(1-x)q^2-\ell^2)^3}\stackrel{n\rightarrow 4}{=}  \frac i{32\pi^2} \label{C3}
\ee
and
\be 
\int\frac {d^4p}{(2\pi)^4}
\int \frac {d^{n-4}\ell}{(2\pi)^{n-4}}  \frac {p^2\ell^2}{(p^2+x(1-x)q^2-\ell^2)^3}\stackrel{n\rightarrow 4}{=}  \frac i{16\pi^2} \,2y\, (1-x-y)\, k_1\!\cdot\! k_2\label{C4}.
\ee
Integration over $x$ and $y$ is elementary and one gets (\ref{T4}). Both RHS's are obtained
by Wick-rotating all the momenta.
 
\section{Local expression of the trace anomaly}
\label{sec:D}

The partition function depending on a classical external source $j_{\mu\nu}$ is
\be 
Z[j_{\mu\nu}]&=& \langle 0| T\,e^{i\int dx \sqrt{|g(x)|} T^{\mu\nu}(x)j_{\mu\nu}(x)}|0\rangle=
e^{iW[j_{\mu\nu}]}\label{Zg}\0\\
&=& \sum_{n=0}^\infty \frac {i^n}{n!} \int  \prod_{i=1}^n dx_i \, \sqrt{|g(x_i)|}\, j_{\mu_i\nu_i}(x_i)\, \langle 0|T\,T^{\mu_1\nu_1}(x_1)\ldots T^{\mu_n\nu_n}(x_n)|0\rangle.
\ee
The generating functional of connected Green functions is
\be 
W[j_{\mu\nu}]= \sum_{n=1}^\infty \frac {i^{n+1}}{n!} \int \prod_{i=1}^n dx_i\,\sqrt{|g(x_i)|}\, j_{\mu_i\nu_i}(x_i)\, \langle 0|T\,T^{\mu_1\nu_1}(x_1)\ldots T^{\mu_n\nu_n}(x_n)|0\rangle_c\label{Wg}.
\ee
We will denote by 
\be 
\langle T^{\mu\nu}(x)\rangle = - \frac 2{\sqrt{|g|}}\frac {\delta W[j]}{\delta j_{\mu\nu}(x)}\Big{\vert}_{j_{\mu\nu}=g_{\mu\nu}}\label{qem}
\ee
the `quantum e.m. tensor'. Let us take now the variation of $W$ with respect to a conformal transformation $j_{\mu\nu}=g_{\mu\nu}$ and $\delta g_{\mu\nu}(x) = 2 \omega(x) g_{\mu\nu}(x)$
where $g_{\mu\nu}= \delta_{\mu\nu}+ h_{\mu\nu}+\cdots$ is a classical metric configuration and $h_{\mu\nu}$ is the field attached to the external legs of the Feynman diagrams of section 3. Due to the arbitrariness of $\omega$, invariance of $W[g]$ under conformal transformations means vanishing the `quantum' e.m. trace,
\be 
\langle T^{\mu}_{\mu}(x)\rangle= 2\sum_{n=1}^\infty \frac {i^{n+1}}{(n-1)!} 
\int \prod_{i=2}^n dx_i\,\sqrt{|g(x^i)|}\, g_{\mu_i\nu_i}(x_i) \,\langle 0|T\,T^{\mu}_{\mu}(x)\ldots T^{\mu_n\nu_n}(x_n)|0\rangle_c\label{Tmumu}.
\ee
This means in particular that all the Green functions $\langle 0|T\,T^{\mu}_{\mu}(x)\ldots T^{\mu_n\nu_n}(x_n)|0\rangle_c$ must vanish in order to guarantee quantum conformal invariance.
In this paper we focus on the amplitude
\be 
\langle T_{\sigma}^{\sigma} (q) \,T_{\mu\nu}(k_1)  \,T_{\mu'\nu'}(k_2)\rangle=
\int d^4x\,d^4y\, d^4z \, e^{i(k_1x+k_2y-qz)}\langle T_{\sigma}^\sigma (z) \,T_{\mu\nu}(x)  \,T_{\mu'\nu'}(y)\rangle\label{3T}
\ee
at one-loop order.
On the basis of the previous discussion, the local expression of the anomaly is obtained by Fourier-antitransforming  (\ref{Ttot}) and inserting it into (\ref{Tmumu}), and, simultaneously, identifying $j_{\mu\nu}(x)=g_{\mu\nu}(x)$, where $g_{\mu\nu}(x)$ satisfies the eom and the de Donder gauge (see section \ref{ssec:3.2}).
One relevant contribution to (\ref{Tmumu}) is $t_{\mu\nu\mu'\nu'\lambda\rho}\,k_1^\lambda k_2^\rho \,\, k_1\cdot k_2 \,\delta(q-k_1-k_2)$, from which
\begin{multline} 
t_{\mu\nu\mu'\nu'\lambda\rho}\int \frac{d^4k_1}{(2\pi)^4}\frac{d^4k_2}{(2\pi)^4} \frac{d^4q}{(2\pi)^4} \,e^{-i(k_1x+k_2y-qz)}  \,k_1^\lambda k_2^\rho \,\, k_1\cdot k_2 \,\delta(q-k_1-k_2)\label{T1xyz}\\
= t_{\mu\nu\mu'\nu'\lambda\rho}\del_x^\lambda \del_x^\tau \delta(x-z)\, \del_y^\rho\del_{y\tau} \delta(y-z)
\end{multline}
Inserting this into (\ref{Tmumu}) we get 
\be 
\langle T^{\mu}_{\mu}(x)\rangle^{(1)}&=&  t_{\mu\nu\mu'\nu'\lambda\rho}\int d^4x d^4y 
(\delta_{\mu\nu}+ h_{\mu\nu})(\delta_{\mu'\nu'}+h_{\mu'\nu'}) \del_x^\lambda \del_x^\tau \delta(x-z)\, \del_y^\rho\del_{y\tau} \delta(y-z)\0\\
&=& 4 \,\epsilon_{\nu\nu'\lambda\rho} \del^\lambda\del^\tau h^{\mu\nu} \,\del^\rho\del_\tau h_\mu^{\nu'}\label{Tmumu1}
\ee
Another relevant contribution is given by (it comes from the term containing $t^{(21)}$)
\begin{multline} 
k_{2\nu}k_{1\nu'}\,\epsilon_{\mu\mu'\lambda\rho}\,k_1^\lambda k_2^\rho \,\,\delta(q-k_1-k_2)\label{T2xyz}\\
= \epsilon_{\mu\mu'\lambda\rho}\int d^4xd^4yd^4z \,e^{i(k_1x+k_2y-qz)}\, \del_x^\lambda \del_{x\nu'}\delta(x-z) \,
\del_y^\rho\del_{y\nu}\delta(y-z)
\end{multline}
Inserting it into (\ref{Tmumu}) we get
\be 
\langle T^{\mu}_{\mu}(x)\rangle^{(2)}= 4\, \epsilon_{\mu\mu'\lambda\rho}\,
\del^\lambda\del_{\tau}h^{\mu\nu}\, \del^\rho\del_\nu h^{\mu'\tau}\label{Tmumu2}
\ee
This result is still Euclidean.


\end{document}